\documentclass[aps,twocolumn,showpacs,preprintnumbers,amsmath]{revtex4}
\usepackage{graphicx}
\usepackage{epsfig}
\usepackage{amsfonts}
\usepackage{amsmath,amssymb}

\begin{document}
\title{Exactly-solvable models for atom-molecule hamiltonians}
\author{J. Dukelsky$^1$, G. G. Dussel$^2$, C. Esebbag$^3$ and S. Pittel$^4$}
\address{$^1$ Instituto de Estructura de la Materia, CSIC, Serrano 123, 28006 Madrid, Spain.}

\address{$^2$ Departamento de F\'{\i}sica Juan Jose
Giambiagi, Universidad de Buenos Aires, 1428 Buenos Aires, Argentina.}

\address{$^3$ Departamento de Matem\'aticas, Universidad de Alcal\'a, 28871
Alcal\'a de Henares, Spain.}
\address{$^4$ Bartol Research Institute, University of Delaware, Newark,
Delaware 19716, USA.}


\begin{abstract}
We present a family of exactly-solvable generalizations of the
Jaynes-Cummings model involving the interaction of an ensemble of
SU(2) or SU(1,1) quasi-spins with a single boson field. They are
obtained from the trigonometric Richardson-Gaudin models by
replacing one of the SU(2) or SU(1,1) degrees of freedom by an
ideal boson. Application to a system of bosonic atoms and
molecules is reported.
\end{abstract}
\maketitle

The Jaynes-Cummings (JC) model \cite{JC} provides a simple
description of the interaction of matter with a radiation field.
It treats a two-level atom in terms of the spin-1/2 generators of
the SU(2) algebra and describes its coupling to a single radiation
field in the so-called Rotating Wave Approximation. Despite its
simplicity, the model has had enormous success in quantum optics,
finding realization in experiments with Rydberg atoms in microwave
cavities \cite{Re} and optical cavities \cite{Ra}.

There have also been several extensions of the JC model that have
likewise proven useful. One example is the Tavis-Cummings (TC)
model \cite{TC}, in which the spin-1/2 operators are replaced by
operators for arbitrary spin, permitting the description of a
collection of equivalent two-level atoms with a radiation field.
This model has been solved exactly using the Quantum Inverse
Scattering Method for and arbitrary SU(2) spin \cite{QISM}.
Another example is the Buck-Sukumar (BS) model \cite{BS}, in which
a specific non-linear interaction between the atoms and the
radiation field is included and which is also exactly solvable.
More general non-linear terms have also been discussed, but they
can only be treated approximately \cite{Vadeiko}. A third example
is an exactly-solvable atom-molecule hamiltonian that describes
the photoassociation of pairs of condensed bosonic atoms -- based
on the algebra SU(1,1), rather than SU(2) -- into molecules with a
linear interaction \cite{Links}, or with a non-linear interaction
\cite{Ryb}.

Similar physics is also at play when a molecular Bose-Einstein
condensate is produced through photoassociation (with or without
the interaction with a Feshbach resonance) in dilute fermion
\cite{Fermion} or boson \cite{Boson} gases. The production of
degenerate bosonic Sodium atom-molecule mixtures has been recently
reported \cite {Muk}, though it is still an open question as to
whether the molecules formed a BEC. Mixtures of fermion atoms and
molecular dimers are better candidates for constructing a
molecular BEC due to the suppression of molecular decay by Pauli
blocking. Indeed, two groups have reported the manufacture of
molecular BECs from $^{40}Ka$ \cite{Gre} and $^{6}Li$ \cite{Jo}
fermionic atoms, respectively. Unfortunately, no current
exactly-solvable model can describe these physical processes.

In this letter, we show how to generalize the JC model to accommodate these physical scenarios, as
well as others, in the context of exactly-solvable models. The extension is to a family of models
that involve an ensemble of SU(2) or SU(1,1) quasi-spins and a single bosonic mode. For the purposes
of our discussion, the SU(2) models describe fermion pairs and their coupling to a bosonic mode,
whereas the SU(1,1) models describe the corresponding physics of bosonic pairs. The SU(2) models
could also be used to describe two-level atoms and a bosonic mode, but we will not discuss such
models here. The generalizations we will describe build on the recently-proposed Richardson-Gaudin
(RG) integrable models \cite{duke1} (for a recent review see \cite{DPS}). Following the presentation
of the models, we will discuss their specific application to a mixture of bosonic atoms and
molecular dimers.

We begin by introducing the generators of the SU(2) and SU(1,1)
algebras, $K_i^0$, $K_i^+$ and $K_i^-=(K_i^+)^{\dag}$, which
satisfy the commutation relations

\begin{eqnarray}
\left[ K_{i}^{0},K_{j}^{+ }\right] = \delta _{ij}K_{i}^{+}~ , ~
\left[ K_{i}^{+},K_{j}^{-}\right] =\mp 2\delta _{ij}K_{i}^{0 } ~.
\label{KS}
\end{eqnarray}

\noindent The upper sign refers to the bosonic SU(1,1) algebra and
the lower sign to the fermionic SU(2) algebra, as they will
throughout this presentation.

In the quasi-spin or pair representation of the SU(2) and SU(1,1)
algebras, the generators are realized in terms of particle
creation and annihilation operators  as

\begin{equation}
K_{j}^{0}=\frac{1}{2}\sum_{m}a_{jm}^{\dagger }a_{jm}\pm \frac{\Omega _{j}}{4}%
\text{ \ ,}~~K_{j}^{+}=\frac{1}{2}\sum_{m}a_{jm}^{\dagger }a_{j\bar{m}%
}^{\dagger }\text{ \ } ~. \label{gen}
\end{equation}
\noindent  Here $a_{jm}^{\dagger }\left( a_{jm}\right) $ creates
(annihilates) a boson or a fermion in the state $|jm\rangle$,
$|j\bar{m}\rangle$ is the state obtained by acting with the time
reversal operator on $|jm\rangle$, and $\Omega_j$ is the total
degeneracy of single-particle level $j$.

There are three families of fully integrable and exactly-solvable
RG models that derive from the SU(2) and SU(1,1) algebras, the
rational, trigonometric and hyperbolic models, respectively
\cite{duke1}. For all three families, it has been shown how to
write the complete set of commuting operators (the integrals of
motion) and the corresponding eigenvalues and eigenvectors. Here
we focus on the trigonometric family, for which the integrals of
motion can be expressed in terms of the generators as

\begin{eqnarray}
R_{i} &=&K_{i}^{0}+2g\sum_{j\left( \neq i\right) }\{\frac{1}{2\sin
(\eta
_{i}-\eta _{j})}[K_{i}^{+}K_{j}^{-}+K_{i}^{-}K_{j}^{+}]  \nonumber \\
&~&\qquad~~~~~~\mp \cot (\eta _{i}-\eta _{j})K_{i}^{0}K_{j}^{0}\}
~. \label{cons}
\end{eqnarray}
\noindent For each degree of freedom $i$, there is one real
arbitrary parameter $\eta_i$ that enters the integrals of motion.
These operators commute amongst themselves and each commutes with
the conserved quantity $K^{0}=\sum_i K^{0}_i$, which is related to
the total number operator.

We now consider the eigenvalue equation for the integrals of
motion, $R_i |\Psi \rangle = r_i |\Psi \rangle$, in the
seniority-zero sector, namely when all particles are paired.
Solutions with broken pairs can also be readily obtained, as in
\cite{duke1}.

In this sector, the eigenstates of $R_i$ are given by
\begin{equation}
\left| \Psi \right\rangle =\prod_{\alpha =1}^{M}B_{\alpha }^{\dagger }\left| 0 \right\rangle , \quad
B_{\alpha }^{\dagger }=\sum_{l} \frac{1}{\sin (e_{\alpha} -\eta_l)} ~K_{l}^{+} ~, \label{ansa}
\end{equation}
\noindent where $|0\rangle$ is a state that is annihilated by all
the $K^-_i$ and $M$ is the number of pairs. The structure of the
collective operators $B^{\dagger}_{\alpha}$ is determined by a set
of $M$ parameters $e_{\alpha }$, which satisfy the set of coupled
nonlinear equations

\begin{equation}
1-\frac{g}{2}\sum_{j}\Omega _{j}\cot (e_{\alpha }-\eta _{j})\pm
2g\sum_{\beta \left( \neq \alpha \right) }\cot (e_{\beta
}-e_{\alpha })=0 ~.\label{Rich}
\end{equation}

\noindent The associated eigenvalues take the form

\begin{eqnarray}
r_{i}&=&\pm \frac{\Omega _{i}}{4}\left\{
1-\frac{g}{2}\sum_{j\left( \neq i\right) }\Omega _{j}\cot (\eta
_{i}-\eta _{j})  \right.   \nonumber \\
&~&\left. \qquad\pm 2g\sum_{\alpha }\cot (e_{\alpha }-\eta
_{i})\right\}~. \label{eigenvalues}
\end{eqnarray}

The important point to note here is that any hamiltonian that can
be written solely in terms of the integrals of motion $R_i$ is
likewise exactly solvable, with precisely the same eigenvectors as
in (\ref{ansa}) and with eigenvalues that are obtained directly
from those in (\ref{eigenvalues}).

We now discuss how to construct an appropriate subset of the
trigonometric RG models that are of relevance to the quantum
problems described in the introduction, involving the interplay of
a set of SU(2) or SU(1,1) systems with a single bosonic mode. To
do this, we use a trick originally proposed by Gaudin
\cite{Gaudin}, which involves replacing one SU(2) or SU(1,1)
degree of freedom by an ideal boson. For specificity, we  denote
the SU(2) or SU(1,1) degree of freedom to be bosonized as $i=0$
and the remaining as $i=1,\dots,L$. In the limit $\Omega_{0}
\rightarrow \infty$, the generators map onto ideal bosons
according to

\begin{equation}
K_{0}^{0}=b^{\dagger }b\pm \frac{\Omega _{0}}{4}~~,\quad K_{0}^{\dagger }=\sqrt{%
\frac{\Omega _{0}}{2}} ~ b^{\dagger } ~. \label{Bos}
\end{equation}

We now introduce a change of notation for the trigonometric
functions that appear in (\ref{cons}) for the selected degree of
freedom, $w_{j}=1/\sin (\eta _{0}-\eta _{j})$~, $v_{j}=\cot (\eta
_{0}-\eta _{j})$ ~, with $w_{l}^{2}-v_{l}^{2}=1$.  Moreover, we
expand these amplitudes in the inverse of the divergent degeneracy
$\Omega _{0}$,

\begin{equation}
w_{l}^2=1+\frac{\varepsilon _{l}}{\Omega _{0}}~,\quad \ v_{j}=-\sqrt{\frac{2}{%
\Omega _{0}}}~\varepsilon _{j} ~, \label{Amp}
\end{equation}
thereby introducing a new set of parameters $\varepsilon_l$ $(l=1,...,L)$ to replace the $\eta_l$'s.

Inserting (\ref{Bos}) and (\ref{Amp}) into (\ref{cons}), we obtain
new integrals of motion that involve the ideal boson degree of
freedom,

\begin{equation}
R_{0}=b^{\dagger }b+G \left[ \sum_{j}\left( b^{\dagger
}K_{j}^{-}+K_{j}^{\dagger }b\right) +\sum_{j}\varepsilon
_{j}K_{j}^{0}\right] , \label{R0}
\end{equation}

\begin{eqnarray}
R_{j}&=&K_{j}^{0}+G \left[ \sum_{i\left( \neq
j\right) }\{\frac{1}{\left( \varepsilon _{i}-\varepsilon _{j}\right) }%
[K_{i}^{+}K_{j}^{-}+K_{i}^{-}K_{j}^{+}] \right. \nonumber \\
&~&\left. \mp \frac{2}{\varepsilon
_{i}-\varepsilon _{j}}K_{i}^{0}K_{j}^{0}\}-[K_{j}^{+}b+K_{j}^{-}b^{+}]-%
\varepsilon _{j}K_{j}^{0}\right] ~, \label{Rj}
\end{eqnarray}
where $G=g\sqrt{\frac{\Omega _{0}}{2}}$.

After bosonization of the selected degree of freedom, the resulting $R_i$ still satisfy the
conditions for an integrable model. They remain hermitian, global, independent, and mutually commute
with one another, thereby constituting a complete set of integrals of motion.  Thus, any hamiltonian
that can be written in terms of these $R_i$ likewise defines an exactly-solvable model.

It is important to note that the set of integrals of motion given
in Eqs. (\ref{R0}) and (\ref{Rj}) define {\it a totally new set of
exactly-solvable models}, even though they were derived from the
trigonometric family of RG models. That they are not simply the
trigonometric family rewritten can be seen by focussing on Eq.
(\ref{Rj}), which gives the form of the new $R_j$ integrals of
motion.  They are in fact identical to those of the rational
family of ESMs, {\it except for the last two terms} which are
essential for ensuring the commutation with the new bosonic
integral of motion $R_0$. In this sense, the new class of models
that we derived can be viewed as an extension of the rational
family to include an extra boson degree of freedom.

Note further that the operator that counts the {\em total} number
of pairs, $M= b^{\dagger} b +\frac{1}{2} \sum_{jm}
a^{\dagger}_{jm} a_{jm}$, also commutes with all $R_i$ and thus
defines a conserved quantity.

Before continuing our derivation of the properties of the exact
solutions associated with this complete set of integrals of
motion, we first write down some of the possible hamiltonians that
could be treated exactly in this way. One that is especially
interesting is obtained directly from the selected integral of
motion $R_0$,
\begin{eqnarray}
H&=&\omega R_{0} \mp \frac{\omega G}{4} \sum_{j} {\Omega_j
\varepsilon_j} \nonumber
\\
&=&\omega b^{\dagger }b+\sum_{jm}\epsilon _{j}~ a^{\dagger}_{jm}
a_{jm}+V\sum_{j}\left( b^{\dagger }K_{j}^{-}+K_{j}^{\dagger
}b\right) ~,\label{GTC}
\end{eqnarray}
where $V=\omega g \sqrt{\frac{\Omega_0}{2}}$ and $\epsilon _{j}=V
\varepsilon _{j}/2$. In the pair representation of SU(2) or
SU(1,1), this hamiltonian describes the interaction of fermionic
or bosonic atom pairs with a diatomic molecule. In the two-level
representation, it generalizes the Tavis-Cummings model to
multi-atoms.

Many other exactly-solvable models can also be constructed in this
way. For example, by taking a linear combination of the other
integrals of motion $R_j$ ($j=1,...,L$) with coefficients
$\varepsilon_j$, we obtain a hamiltonian of the form

\begin{eqnarray}
H &=& \sum_{j}\varepsilon _{j}\left( 1-G\varepsilon _{j}\right)
K_{j}^{0}-G\left[ \sum_{i\neq
j}\{\frac{1}{2}[K_{i}^{+}K_{j}^{-}+K_{i}^{-}K_{j}^{+}] \right.
\nonumber \\
&~& ~~~~ \left. \mp
K_{i}^{0}K_{j}^{0}\}+\sum_{j}\varepsilon _{j}[K_{j}^{+}b+K_{j}^{-}b^{+}]%
\right] ~. \label{HGen}
\end{eqnarray}

\noindent Included are those that contain pairing interactions and
those with level-dependent atom-molecule couplings.

We now discuss how to rewrite the seniority-zero solutions for the
trigonometric RG models to apply when one of its degrees of
freedom has been replaced by an ideal boson in the
infinite--$\Omega_0$ limit. Defining $x_{\alpha
}=\sqrt{\frac{\Omega _{0}}{2}}\cot \left( e_{\alpha }-\eta
_{0}\right) $, the Richardson-Gaudin equations (\ref{Rich}) that
define the parameters $e_{\alpha}$, and thus the $x_{\alpha}$,
become

\begin{equation}
\frac{1}{2 G}-\frac{1}{2} x_{\alpha }-\frac{1}{4}\sum_{j}\frac{\Omega _{j}}{\varepsilon
_{j}-x_{\alpha }}\mp \sum_{\beta \left( \neq \alpha \right) }\frac{1}{x_{\beta }-x_{\alpha }}=0 ~.
\label{Richa}
\end{equation}

The corresponding expressions for the eigenvalues
(\ref{eigenvalues}) associated with the new integrals of motion
(\ref{R0}-\ref {Rj}) are

\begin{equation}
r_{0}=\pm \frac{G}{4} \sum_j \Omega _{j}~ \varepsilon_j +G
\sum_{\alpha} x_{\alpha}
 ~,\label{r0}
\end{equation}

\begin{eqnarray}
r_{i}= \pm \frac{\Omega _{i}}{4} &&\left\{ 1\mp 2  G\left[
\frac{1}{2}\varepsilon _{i}\pm \frac{1}{4}\sum_{j(\neq
i)}\frac{\Omega _{j}}{\varepsilon _{i}-\varepsilon _{j}} \right.
\right. \nonumber \\
 &&\left. \left. ~~~~~~~~~~ +\sum_{\alpha
}\frac{1}{x_{\alpha }-\varepsilon _{i}} \right] \right\}
~,\label{ri}
\end{eqnarray}
\noindent while the seniority-zero eigenvectors take the form
\begin{equation}
\left| \Psi \right\rangle =\prod_{\alpha =1}^{M}\left( b^{\dagger }+\sum_{l}%
\frac{1}{x_{\alpha }-\varepsilon _{l}}K_{l}^{+}\right) \left|
0\right\rangle ~. \label{Eig}
\end{equation}
We note here that each independent solution of the set on
non-linear coupled equations (\ref{Richa}) defines an eigenstate
(\ref{Eig}) that is common to the $L+1$ integrals of motion
(\ref{R0}-\ref {Rj}) and has eigenvalues (\ref{r0}-\ref {ri}).

The eigenvalues of the hamiltonian (\ref{GTC}), for example, can
be obtained from the eigenvalues (\ref{r0}) of $r_0$  as
\begin{equation}
E=V\sum_{\alpha} x_{\alpha} ~.\label{E}
\end{equation}
\noindent The eigenvalues of the hamiltonian (\ref{HGen}) can
similarly be obtained from the eigenvalues (\ref{ri}) of the $R_i$
operators.

It is worth noting here that the solutions given in
(\ref{Richa}-\ref{Eig}) are identical to those of the
Tavis-Cummings model \cite{QISM} for a single SU(2) spin and to
those for the atom-molecule model of \cite{Links} for a single
SU(1,1) bosonic level.

Important observables in these models are the occupation
probabilities of the various degrees of freedom. They can be
obtained from the integrals of motion using the Hellman-Feynman
theorem, viz:
\begin{equation}
\left\langle K_{i}^{0}\right\rangle  =\left\langle R_{i}\right\rangle
-G\left\langle \frac{\partial R_{i}}{\partial G}\right\rangle =r_{i}-G\frac{%
\partial r_{i}}{\partial G} ~,
\end{equation}
from which we obtain for the occupations numbers

\begin{equation}
n_i=-\Omega_i G^2 \sum_{\alpha }\frac{1}{\left(
x_{\alpha }-\varepsilon _{i}\right) ^{2}}\frac{\partial x_{\alpha }}{%
\partial G} ~.
\end{equation}
The derivatives of the $%
x_{\alpha }$ are obtained by differentiating the RG equations
(\ref{Richa}) with respect to $G$, which gives

\begin{eqnarray}
&&\left[ \pm \frac{1}{2} \pm \frac{1}{4}\sum_{j}\frac{\Omega_{j}}{\left( x_{\alpha } -\varepsilon
_{j}\right) ^{2}}+\sum_{\beta \left( \neq \alpha \right) }\frac{1}{\left( x_{\beta }-x_{\alpha
}\right) ^{2}}\right]
\frac{\partial x_{\alpha }}{\partial G} \nonumber \\
&&~~~~~~~~-\sum_{\beta \left( \neq \alpha \right) }
\frac{1}{\left( x_{\beta }-x_{\alpha }\right) ^{2}}\frac{\partial x_{\beta }%
}{\partial G} = \mp \frac{1}{2G^{2}} ~. \label{ocu}
\end{eqnarray}

\begin{figure}[htbp]
\hspace{-1cm}
\includegraphics[width=5cm]{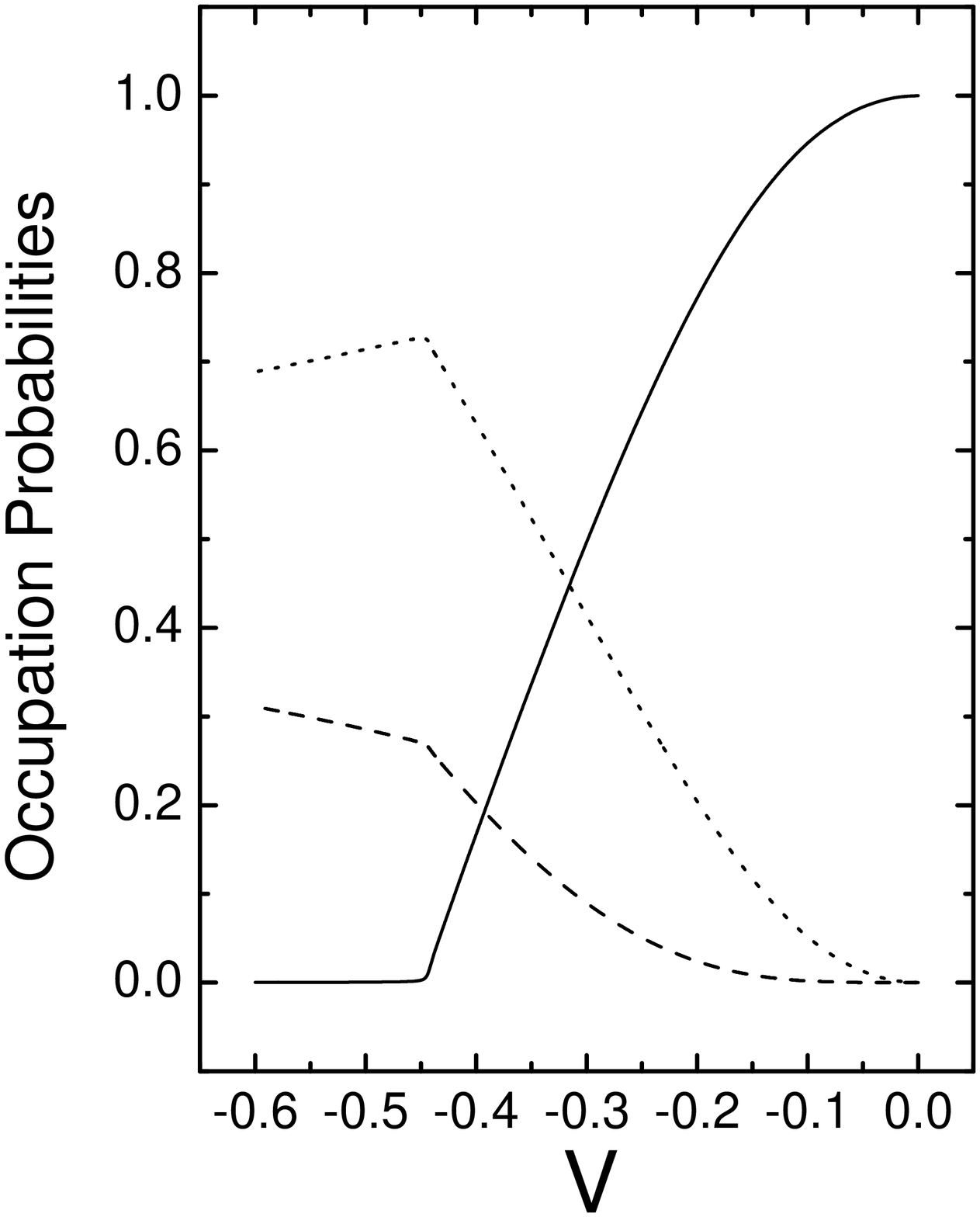}
\vspace{-0.5cm} \caption{Occupation probabilities as a function of
the interaction $V$ for a molecule energy $\omega=10$. The solid
line corresponds to condensed atoms, the dashed line to
non-condensed atoms, and the dotted line to condensed molecules.}
\label{fig1}
\end{figure}

We now turn to a specific application of these new
exactly-solvable models as an illustration of the procedure to
solve eq. (\ref{Richa}) and (\ref{ocu}). We consider a mixture of
bosonic atoms confined to a 3D isotropic trap coupled to a
molecular two-particle bound state, a molecular dimer, and model
it through the hamiltonian (\ref{GTC}). This hamiltonian does not
contain an atom-atom interaction, which could be included by using
the more general hamiltonian of (\ref{HGen}). In the hamiltonian
we use, $\omega$ is the energy of the molecular dimer above that
of the Feshbach resonance and is the negative of the detuning
parameter. Also, $V$ is the atom-molecule interaction strength,
$\epsilon_j=j$ ($j=0,1,\dots$) are the single-atom energies in a
$3D$ isotropic trap, and $\Omega_j=(j+1)(j+2)/2$ are the level
degeneracies. The phase diagram and dynamics of this model have
been studied in several recent works, {\em e.g.} \cite{Ja}. To
make contact with ref. \cite{Ja}, our detuning parameter is
related to theirs by $\omega=\delta$ and our atom-molecule
coupling is $V=-\frac{K}{2 \sqrt{M}}~$.

As noted before, the complete set of seniority-zero eigenstates
arise from different solutions of the equations (\ref{Richa}). For
boson systems with negative couplings $V$, the parameters
$x_{\alpha}$ for the ground-state solution are real and positive.
Excited states correspond to different partitions of the
$x_{\alpha}$ in intervals defined by the $\varepsilon_j$.

We have performed calculations for a system with $M=500$ pairs and
two values of the molecular energy, $\omega=10$ and $\omega=-10$,
as a function of the negative coupling $V$. The atom space was
truncated to $L=50$ harmonic oscillator shells. In Fig.\ref{fig1},
we show the occupation probabilities of the atomic condensate
(solid line) and the atomic depletion (dashed line), and the
fraction of molecules (dotted line) as a function of $V$ for
positive molecular energy ($\omega = 10$). As can be seen, a
quantum phase transition takes place at $V\simeq -0.45$.
Interestingly, the occupation of the atom condensed state is
negligible for $V < -0.45$ and the atomic fraction is distributed
among all harmonic oscillator levels. A pure molecular state does
not exist for any value of $V$. This quantum phase transition was
recently studied using mean-field and renormalization group
techniques and it was concluded that it lies in the Ising
universality class \cite{QPT}. The exact solution offers a unique
opportunity to unravel the critical properties around the
transition point.

\begin{figure}[htbp]
\hspace{-1cm}
\includegraphics[width=5cm]{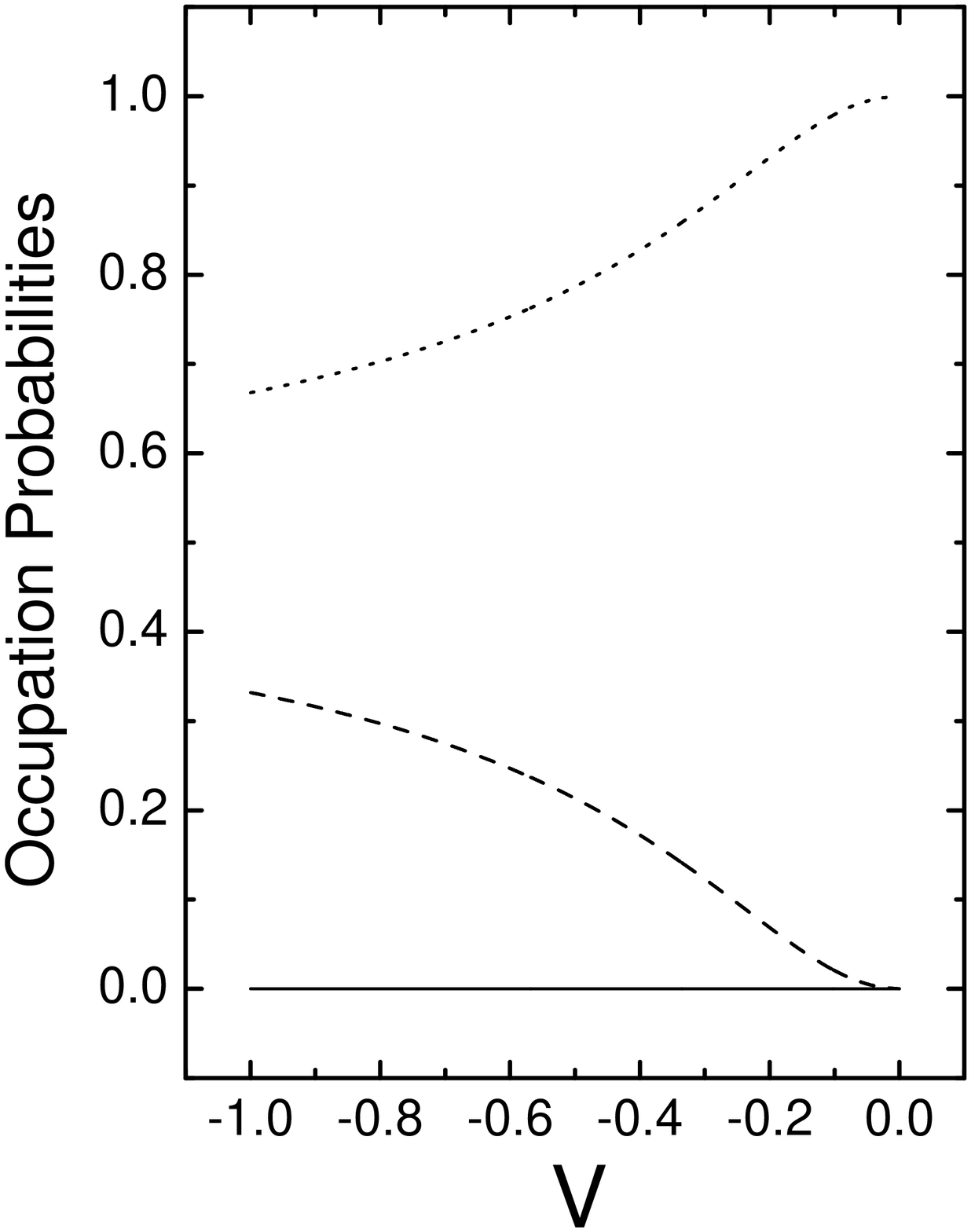}
\vspace{-0.5cm} \caption{Occupation probabilities as a function of
the interaction $V$
 for a molecule energy $\omega=-10$. The solid line corresponds to condensed atoms,
 the dashed line to non-condensed atoms, and the dotted line to condensed molecules.}
\label{fig2}
\end{figure}
In Figure \ref{fig2} we show results for negative molecular
energy, $\omega=-10$. The system is purely molecular for weak
coupling. Molecules begin to decay to pair atomic states, as the
interaction strengthens, but there is no phase transition and the
occupation of the lowest trap level is always negligible. A
detailed study of the phase diagram that emerges from the exact
solutions will be given elsewhere.

In closing, we have presented in this paper a new family of
integrable models for atom-molecule systems. The models are
exactly solvable for fermionic and bosonic atoms interacting with
molecular dimers. There is a large freedom to select the
parameters of the hamiltonian, allowing for the description of
quite general realistic systems. We have presented initial results
for a mixed system of trapped bosonic atoms and molecular dimers.
Application to systems of fermionic atoms and molecular dimers is
of special interest due to the recent generation of an ultra cold
molecular BEC from the conversion of $^{40}Ka$ \cite{Gre} or
$^{6}Li$ \cite{Jo} fermionic atoms. Such models can also be used
to explore the BCS to BEC crossover, from a condensate dominated
by Cooper pairs to a condensate dominated by molecular dimers.
Finally, it is possible to use alternative realizations of these
models to describe problems of importance in quantum optics and
perhaps elsewhere.

After this work has been submitted, we learn about a recent
preprint on generalized integrable matter-radiation models
\cite{Kun}. The extended the Jaynes-Cummings Hamiltonians treated
there have non-hermitian atomic interactions.

This work was supported by the Spanish DGI under grant \#s
BFM2003-05316-C02-01/02, by the US National Science Foundation
under grant \#s PHY-9970749 and PHY-0140036, and by UBACYT X-204.
Fruitful discussions with L. L. Sanchez-Soto are acknowledged.


\begin{references}

\bibitem{JC} E. Jaynes and F. Cummings, Proc. IEEE {\bf 51}, 89 (1963).

\bibitem{Re} G. Rempe, H. Walther, and N. Klein, Phys. Rev. Lett. {\bf 58}, 353 (1987).

\bibitem{Ra} M.G. Raizen {\it et al.}, Phys. Rev. Lett. {\bf 63}, 240 (1989).


\bibitem{TC} M. Tavis, and F.W. Cummings, Phys. Rev. B {\bf 170}, 379 (1968)


\bibitem{QISM} N.M. Bogoliubov, R.K. Bullough, and J. Timonen, J. Phys. A {\bf 29}, 6305 (1996).

\bibitem{Ryb} A. Rybin, {\it et al.} J. Phys. A {\bf 31}, 4705 (1998).


\bibitem{BS} B. Buck, and C.V. Sukumar, Phys. Lett A {\bf 81}, 132 (1981).

\bibitem{Vadeiko} I.P. Vadeiko, G.P. Miroshnichenko, A.V. Rybin, and J. Timonen, Phys. Rev. A {\bf 67},
053808 (2003).

\bibitem{Links} J. Links, H.-Q. Zhou, R.H. McKenzie, M.D. Gould, J. Phys. A {\bf 36}, R63 (2003).

\bibitem{Fermion} J.N. Milstein, S.J.J.M.F. Kokkelmans, and M.J. Holland, Phys. Rev. A {\bf}66, 043604 (2002).

\bibitem{Boson} K. M{\o}lmer, Phys. Rev. Lett. {\bf 90}, 110403 (2003).

\bibitem{Muk} K. Xu, {\it et. al.}, Phys. Rev. Lett. {\bf 91}, 210402 (2003)

\bibitem{Gre} M. Greiner, C. A. Regal, and D. S. Jin, Nature {\bf 426}, 537 (2003).

\bibitem{Jo} S. Jochim, {\it et. al.}, Science {\bf 302}, 2101 (2003).

\bibitem{duke1} J. Dukelsky, C. Esebbag, and P. Schuck, Phys. Rev. Lett. {\bf 87}, 066403 (2001).

\bibitem{DPS}  J. Dukelsky, S. Pittel and G. Sierra, submitted to Reviews of Modern Physics (2003).


\bibitem{Gaudin} M. Gaudin, J. Phys. (Paris) {\bf 37}, 1087 (1976).

\bibitem{Ja} M. Kostrun, and J. Javanainen, cond-mat/0308259.

\bibitem{QPT} L. Radzihovsky, and J. Park, cond-mat/0312237; M.W.J.
Romans {\it et al.}, cond-mat/0312446.

\bibitem{Kun} A. Kundu, quant-ph/0307102.
\end{references}
\end{document}